\documentstyle[eqsecnum,aps,epsf]{revtex}

\input epsf
\begin{document}

\title{ Non-Thermal Production of WIMPs and 
the Sub-Galactic Structure of the Universe}
\author{W.B. Lin~$^{1}$,  D.H. Huang~$^{2,1}$,   X. Zhang~$^{1}$,  R. Brandenberger$^{3}$
}
\address{
~\\$^1$
Institue of High Energy Physics, Chinese Academy of Science, P.O. Box 918-4, 
Beijing 100039, P.R. China
~\\$^2$
Deptartment of Physics, Peking University, Beijing 100871, P.R. China
~\\
~$^3$
Physics Department, Brown University, Providence, RI, 02912, USA
}
\maketitle

\begin{abstract}
There is increasing evidence that conventional cold dark matter (CDM) models 
lead to conflicts between
observations and numerical simulations of dark
matter halos on sub-galactic scales. Spergel and Steinhardt showed 
that if the CDM is strongly self-interacting, then the conflicts disappear. However,
the assumption of strong self-interaction would rule out
the favored candidates for
CDM, namely weakly interacting massive particles (WIMPs), such as the neutralino. 
In this paper we propose 
a mechanism of non-thermal 
production of WIMPs and study its implications on the power spectrum.
We find that the non-vanishing velocity of the WIMPs suppresses the
power spectrum on small scales compared to what it obtained in the conventional
CDM model. Our results show that, in this context, WIMPs as candidates for dark matter
can work well both on large scales and on sub-galactic scales. 

\end{abstract}

There is strong evidence for the existence of a substantial amount of cold
dark matter (CDM). The leading candidates for CDM are weakly
interacting massive particles (WIMPs), such as the neutralino. The
neutralino is the lightest supersymmetric particle.
In models with R parity it is stable, and its mass density in the
universe is generally assumed to be a relic of an initially thermal
distribution in the hot early universe. Assuminig, in addition, the presence
of a small 
cosmological constant, the CDM scenario is consistent 
with both the observations of the large scale structure of the universe ($\gg
1$Mpc)
and the fluctuations of the cosmic microwave background\cite{BOPS}. 
Many experiments searching for neutralino dark matter
particles are under way.

The collisionless CDM scenario, however,  
predicts too much power on small scales,
such as a large excess of dwarf galaxies\cite{klypin,moore}, 
the over-concentration of dark matter in dwarf
galaxies\cite{moore94,burkert,MB} 
and in large galaxies\cite{NS}. Recently Spergel and Steinhardt proposed
a new concept of dark matter with strong
self interaction\cite{SS}. 
This puts WIMPs as candidates for dark matter
in considerable jeopardy\cite{WDFMSS,KL}.

In this paper we propose a scenario with non-thermal production of WIMPs.
These WIMPs could be  
relativistic when generated. Their comoving free-streaming 
scales could be as large as of the order 0.1 Mpc or larger. 
The density fluctuations on scales less than the free-streaming scale would then
be severely suppressed. Consequently 
the discrepancies between the observations of dark matter halos on
the sub-galactic scales and the predictions of the standard WIMPs dark matter 
picture could be resolved.

To begin with, we consider a general case of non-thermal production of the
neutralinos by the decay of topological defects such as cosmic string\cite{Robert}, by the 
decay of an unstable heavy particle, or produced non-thermally
by the reheating process in a scenario of inflation at low energy scale\cite{Meng} (see also 
\cite{Kolb}). The momentum distribution  
function of the neutralinos is for simplicity assumed to be Gaussian: 
\begin{equation}
f(p)=\frac{A}{\sqrt{2\pi}\sigma} \exp \left(-\frac{(p-p_{c})^2}{2\sigma^2}\right)~, \label{fp} 
\end{equation}
where $p_{c}$ is the central value and $\sigma$ describes the width of the distribution.

Given a model, the parameters $p_c$ and $\sigma$ can be determined. For
instance, in the supersymmetric version of the $U_{B-L}(1)$ model, the
(Higgsino-like) neutralinos arise
directly from the decay of the right-handed neutrinos and their
superpartners \cite{Robert}. In this model, $p_c$ is about a half of the mass of the mother
particles. For a two-body decay, if the mother particle is at rest, the 
distribution function $f(p)$ is a $\delta$-function. The value of $\sigma$
characterizes the average non-vanishing velocity of the mother particles
(when $\sigma \rightarrow 0$, $f(p)$ approaches a $\delta$-function). In the 
model of Ref.\cite{Robert}, since the mother particles when released from the 
cosmic string loop are non-relativistic, $\sigma$ is small compared with
$p_c$ and one has $p_c\simeq <p> \simeq <p^2>^{1/2}$.

In Eq.(\ref{fp}), $A$ is a normalization factor determined 
by the energy density of the non-thermal component
\begin{equation}
\rho_{NT}=4 \pi \int E(p) f(p) p^2dp~,
\end{equation} 
where $E(p)=(p^2+m^2)^{1/2}$ and $m$ is the rest mass of
the dark matter particle. 
Given that the physical momentum $p(t)$ scales as the inverse of
the cosmic scale factor $a(t)$, we define $r\equiv a(t)p(t)/m$.
During  
cosmic evolution $r$ is a constant.
Throughout this paper we set $a(t_0)=1$, so $r$ can be understood as
the velocity of the particles at the present time
(note that the dark matter particles are
non-relativistic now even though they are relativistic when generated).

The comoving free-streaming scale $R_{f}$ for the non-thermal 
particles can be calculated as follows\cite{KT,BMY}:
\begin{eqnarray} 
\nonumber R_f &=& \int_{t_i}^{t_{EQ}}\frac{v(t')}{a(t')}dt' 
\simeq \int_{0}^{t_{EQ}}\frac{v(t')}{a(t')}dt' \\
& \simeq & 2r_c t_{EQ}(1+z_{EQ})^2 \ln 
\left(\sqrt{1 + \frac{1}{r_c^2(1+z_{EQ})^2}} + \frac{1}{r_c(1+z_{EQ})}\right)~,
\label {Rf}
\end{eqnarray}
where $r_c\equiv a(t)p_c(t)/m=p_c(t_0)/m$ and the subscript '$_{EQ}$' denotes 
radiation-matter equality.
Below the free-streaming scale, the power spectrum will be severely damped. 
To account for the lack of substructure in the Local Group, N-body 
simulations study show that the free-streaming scale of the dark matter 
should be $\sim 0.1 Mpc$\cite{CAV}. This 
corresponds to $r_c\sim 10^{-7}$, which gives rise to a
constraint on the parameters of our model.

To calculate the power spectra for our models, we adapt the CMBFAST code
\cite{SZ} so that it applies to a thermal as well as a non-thermal
dark matter distribution. 
To check the program we replaced the non-thermal distribution 
function in our adapted code by the standard thermal distributions for
hot dark matter and cold dark matter respectively, and calculated
the power spectra, obtaining the same results as with the standard
CMBFAST code.

In the numerical calculations, we took the presently favored values
$\Omega_m=0.4$,~$\Omega_\Lambda=0.6$,~$h=0.65$ and $\Omega_bh^2=0.02$ of
the relevant cosmological parameters, 
where $\Omega_{m}$,~ $\Omega_{\Lambda},$ and $\Omega_{b}$ are the
ratios of the contributions from total mass, 
vacuum energy, and baryons to the total density of the universe.
In general, there are two sources of WIMPs which contribute to $\Omega_m$,
one is from thermal production, the other is from non-thermal
production. For simplicity, in our study  
we neglected the contribution of WIMPs from the thermal production mechanism.

In Fig.1, we show the power spectrum of our model with $r_c=1.5\times
10^{-7}$. For comparison, we also plot the power spectra for the conventional CDM
model and the warm dark matter (WDM) model with $m_W=1$keV 
(For this mass the free-streaming scale is $0.11$Mpc\cite{NSDM}).
Throughout this paper, we normalize all power spectra by the
COBE observations\cite{BW}.

We have varied the parameters of the non-thermally produced WIMPs dark matter
model in the numerical calculations.
Different values of $r_c$ gives rise to different power
spectra. However
the spectrum is independent of the rest mass $m$ 
of the dark matter particle (for fixed $r_c$), and is very insensitive to $\sigma$ 
for $\sigma \ll p_c$.
This can be understand easily from Eq.(\ref{Rf}) 
which shows that
the comoving free-streaming scale $R_f$ only depends on $r_c$. 

One can see from Fig. 1 that on large scales, the three power
spectra are the same, which shows that our model retains all the 
merits of the conventional CDM model
on large scales, while on sub-galactic scales, the power spectrum of 
non-thermal dark matter (NTDM) model is damped
severely relative to that of the CDM model, and is close to 
that of the WDM model with $m_W=1$keV.
Ref.\cite{CAV} argued that such a warm dark matter scenario provides a
solution to the substructure problem.

The model we present in this paper with non-thermal production of WIMPs 
provides a promising scenario for large-scale structure formation 
of the universe. However, we need to check its consistency
with observations on small scales, especially
on scales of the Lyman-$\alpha$ forest.
In Fig.2 we plot the power spectrum of our model and the observed
power spectrum of the Lyman-$\alpha$ system at $z=2.5$ 
shown as the filled circles with error bars\cite{CWPHK,PWCHKP}. 
For comparison we also give the power spectra for the conventional CDM and WDM models. 
In fitting to the observed data,
we choose the primordial spectral index $n=0.97$ for all models. The mass for the WDM particles
is choosen as $750$eV, and the parameters for the NTDM models are $r_c=(1.3,1.4,1.5)\times 10^{-7}$,
respectively. 

To obtain a quantitative constraint on $r_c$, we
closely follow the study by Narayanan et al in Ref.\cite{NSDM}. They
studied explicitly the power spectrum of WDM models and 
came to the conclusion that any mechanism which suppresses the
conventional CDM linear
power spectrum more severely than a 750eV WDM particle will be inconsistent
with Lyman-$\alpha$ forest observations in the scale range
$-2.5\leq \lg [k(km^{-1}s)]\leq -1$. This requires
that the linear theory power spectrum of any substitute to the conventional CDM models 
be higher than that of a WDM model with $m_W=750$eV\cite{NSDM}. 
For a flat universe with $\Omega _{\Lambda}=0.6$ 
the range of these scales corresponds to 0.4$h^{-1}$ Mpc$\leq k\leq$ 12.8
$h^{-1}$ Mpc. From Fig. 2
one can see that the larger the value of $r_c$ is, the lower the small-scale power
spectrum becomes. By comparing the values of the power spectra at the upper limit
of the above range in $k$ with the power spectrum of 
the WDM model with $m_W=750$eV, we obtain an upper limit on $r_c$ of $r_c\leq 1.5\times 10^{-7}$.

Next we examine the constraints on our model by the recent studies
of the phase-space
density\cite{HD,Sellwood,DH}.
The particle phase-space density $Q$ is defined as $Q\equiv \rho/<v^2>^{3/2}$, where $\rho$ is the 
energy density and $<v^2>$ is the mean square value of the particle velocity.
The astronomically observable quantity is the mean coarse-grained phase-space density.
In the absence of dissipation, the coarse-grained phase
space density can only decrease from its primordial value. 
Thus, one can use the observed maximum phase-space density to set 
a lower limit on the phase-space density for the dark matter particles.
The highest observed phase-space density is obtained from
dwarf spheroidal galaxies: 
$Q_{obs}\sim 10^{-4}$M$_{\odot}$pc$^3$(km/s)$^{-3}$~\cite{DH}. 
For our models, $<v^2>\simeq r_c^2$ at the present time, and therefore 
\begin{equation}
Q_0\simeq \rho_{NT0}/r_c^3~.  \label{Q} 
\end{equation}
Because the primordial phase
 space density decreases with time when the particles are 
relativistic and becomes a constant after the particles become
non-relativistic, one requires
$Q_0>Q_{obs}$, which can be translated into a
 constraint on $r_c$:
\begin{equation}
r_c<(\rho_{NT0}/Q_{obs})^{1/3}\sim 2.5\times 10^{-7} \, .
\end{equation} 
This limit is of the same order of magnitude but slightly weaker than that from the observations 
of the Lyman-$\alpha$ forest.

In conclusion, we have demonstrated that our model with non-thermally produced WIMPs
is consistent with observations on scales ranging from 
the Lyman-$\alpha$ systems to the CMB.
Compared to the conventional CDM model, the power spectrum of our model on
sub-galatic scales is severely damped, which will account both for the lack
of substructure in the Local Group and for the observed smooth
inner halos\footnote{
There are other alternatives to the Spergel-Steinhardt solution to the
discrepancies, for instance, the warm dark matter models\cite{CAV,HD,HS,SD},
the proposal of damping of the primordial spectrum on  sub-galactic scales in
some broken-scale-invariance (BSI) inflationary scenarios \cite{KL},
among others\cite{goodman,peebles,RT,HBG,NEF,GS,BKW,BGS,cen,KKT,hannestad}.}.  
In addition, during the structure formation, the non-vanishing velocity of the WIMPs may
further reconcile the discrepancies between the theoretical predictions 
and observations on sub-galactic scales\cite{DH,AFH}. However, the detailed effect of the dark 
matter velocity 
on the formation of halos needs further study with high resolution N-body simulations.  

To solve the sub-galactic-scale problem, our study in this paper requires the value
$r_c\sim 1.5\times 10^{-7}$. This has implications on the models of
non-thermal production of WIMPs. In the scenario of Ref.\cite{Robert}
where neutralinos are released by cosmic string decay,
$r_c$ can be approximately expressed as
$r_c=(T_0/2m)(M/T_d)$, where $T_0$ is the present temperature, $T_d$ 
is the temperature when the string decay first leads to nonthermal
neutralinos, and
$M$ is the characteristic mass of the mother particle and 
the scale of the defect which determine the initial momentum.
For $m$ in the range of $50$Gev to $100$Gev, $M$
is required to be in the range of $1.6\times 10^8$Gev 
to $6.4\times 10^8$Gev. 
This is consistent with the results given in Ref.\cite{Robert}
and shows that the value which we used in the text is quite realistic.

In summary, the discrepancies between theory and observations on 
sub-galactic scales disfavores the conventional WIMPs CDM model. In this
paper we show that 
if the dark matter particles have a 
non-thermal origin,
these discrepancies can be resolved and the WIMPs 
remain good candidates for the dark matter particle.

\acknowledgments

We wish to thank R. Croft and M. Zaldarriaga for helpful correspondence, and
acknowledge M.C. Lee and Z.-H. Lin for discussions.
This work was supported in part by the National Natural Science Foundation of China
and by the U.S. Deparment of Energy under Contract DE-FG02-91ER40688, TASK A.
Some of the work took place during the PIMS/APCTP String Cosmology Workshop (July 24
- Aug. 4, 2000).

\newpage

\begin{figure}  
\epsfxsize=6.0 in \epsfbox{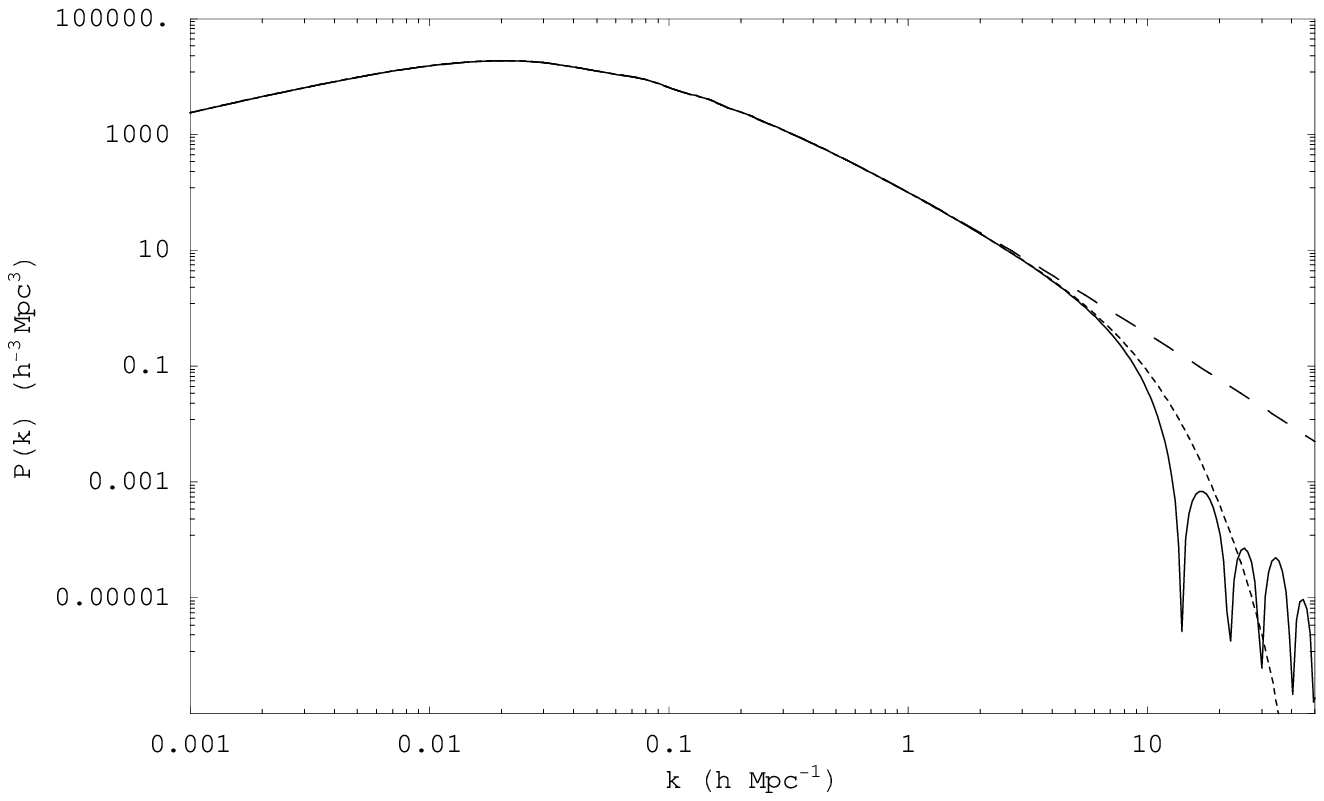}  
\vskip 2mm
\caption{Comparison of the power spectra of the CDM model(long dashed curve), the WDM model
with $m_W=1$keV(short dashed curve) and the NTDM model
with $r_c=1.5\times 10^{-7}$ (solid curve).
}
\end{figure}

\newpage

\begin{figure}  
\epsfxsize=6.0 in \epsfbox{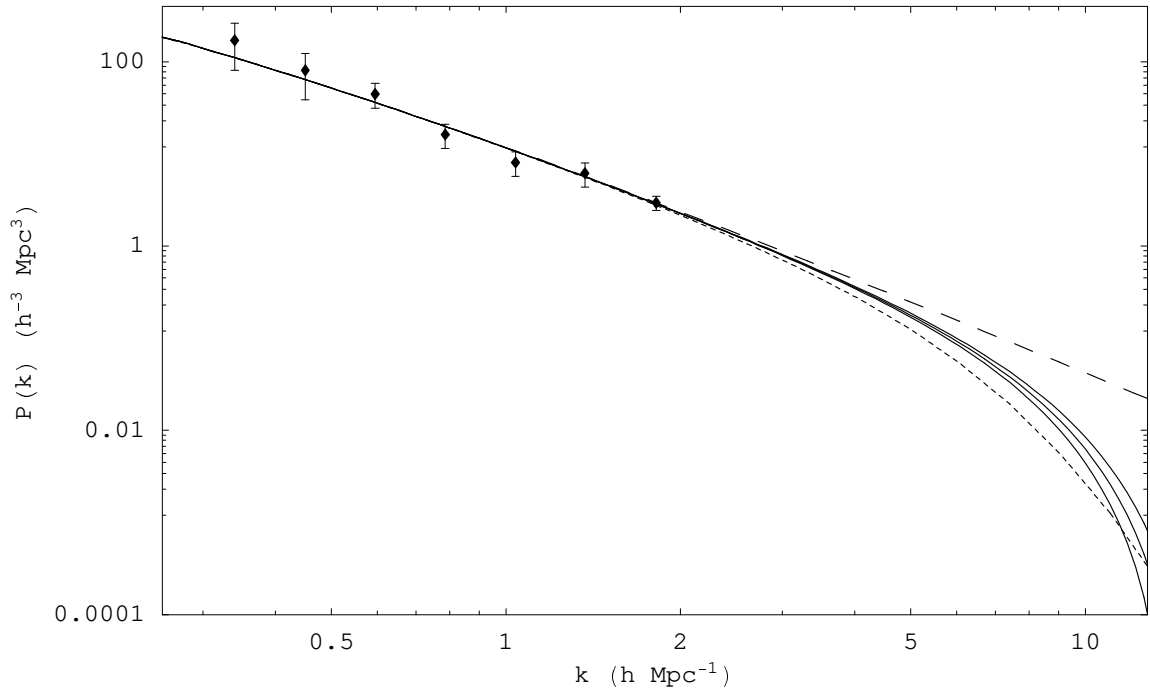}  
\vskip 2mm
\caption{The power spectra of the CDM model (long dashed curve), the WDM model
with $m_W=750$eV (short dashed curve) and the NTDM models 
with $r_c=(1.3,1.4,1.5)\times 10^{-7}$ (solid curves, from top down), 
compared to the observed lyman-$\alpha$ $P(k)$ at $z=2.5$ (filled circles with error bars).
}
\end{figure}

\end{document}